\documentclass[a4paper,english]{article}
\usepackage[T1]{fontenc}
\usepackage[latin1]{inputenc}
\usepackage{babel}
\usepackage{graphics}
\usepackage{epsf,epsfig,psfrag}
\usepackage{amssymb}

\makeatletter

\providecommand{\LyX}{L\kern-.1667em\lower.25em\hbox{Y}\kern-.125emX\@}

\makeatother
\begin{document}

\baselineskip 14pt

\title{\begin{flushright}
{\normalsize IPPP/03/13\\
DCPT/03/26\\}
\end{flushright}\vspace{1cm}
Interactions in Intersecting Brane Models}
  \author{S.A.Abel and A.W.Owen}
 \date{\textit{IPPP, Centre for Particle Theory, Durham
  University, DH1 3LE, Durham, U.K.}}
  
\maketitle

\begin{abstract}
\end{abstract}
We discuss tree level three and four point scattering
amplitudes in type II string models with matter fields localized at the
intersections of D-brane wrapping cycles. Using conformal field theory
techniques we calculate the four fermion amplitudes. These give
"contact" interactions that can lead to flavour changing effects. We show
how in the field theory limit the amplitudes can be interpreted as the
exchange of Kaluza-Klein excitations, string oscillator states and stretched
heavy string modes. 

\section{Introduction}

The discovery of D-branes as non-perturbative objects in any theory of open
and closed strings has led to great progress in our understanding of the
structure of string theory. Furthermore, with the correspondence between
D-branes and gauge theories their status as objects of fundamental importance
to string phenomenology has been well established.

The requirement of chirality has led to a number of scenarios involving
D-branes. These include D-branes on orbifold singularities~\cite{botup} and
more recently intersecting brane models~\cite{ralph1,ibws}, which exploit
the fact that chiral fermions can arise at brane
intersections~\cite{bangles}. The spectrum is then determined by the
intersection numbers of the D-branes which wrap some compact
space. This gives a simple topological explanation of family replication 
which is rather attractive. 

The success of the intersecting brane scenario in producing semi-realistic
models has been well documented, for
example 
refs.\cite{ralph1,chiral,ralph2,ralph4,ralph5,ralph6,smcomp,moresm,bai,hugetitle}.
It is possible to construct models similar to the standard
model~\cite{just,ralph3,just2,exsm2,cvetic3,exsm1} and also models with N=1
supersymmetry~\cite{cvetic1,cvetic2,susy,susyorient}, although this latter possibility is more difficult to achieve.
Recently, attention has been
diverted to a more detailed analysis of the phenomenology of such
models~\cite{deform,gthresholds}. In particular, computation of yukawa couplings~\cite{paper1,qmass,yukawa}
and flavour changing neutral currents~\cite{ams}. This paper details the computation of the three point and
four point functions of string states localised at intersecting branes. These results
allow the estimation of other important constraints on viable configurations of
intersecting branes.

Much of our analysis will be aided by the technology discussed
in~\cite{hamidi,dixon} for closed strings on orbifolds. This is due to an
analogy between twisted closed string states on orbifolds and open strings at
brane intersections. This analogy is the subject of our first section. As a
warm up we
then proceed to a determination of the classical part of the three point function, for which the
computation runs along similar lines to the closed string cases
in~\cite{munoz,het}. The complete calculation of the four point function follows. We
will consider the four point function on only 2 sets of intersecting branes, so that
the classical part is actually somewhat easier to deal with than the three point 
function. The quantum part we evaluate using CFT techniques.

\section{Remarks on closed and open string twisted states}
In order to calculate the three point and four point functions, we will make use of the
analogy between open string states that are stretched between branes at
angles and closed string twisted states on orbifolds. This analogy allows us to
make use of the CFT technology first discussed in ref.\cite{dixon,hamidi}. 
Hence, this first section will be devoted to a description of this 
correspondence.

\begin{figure}
{\centering \resizebox*{0.6\textwidth}{0.3\textheight}{\includegraphics{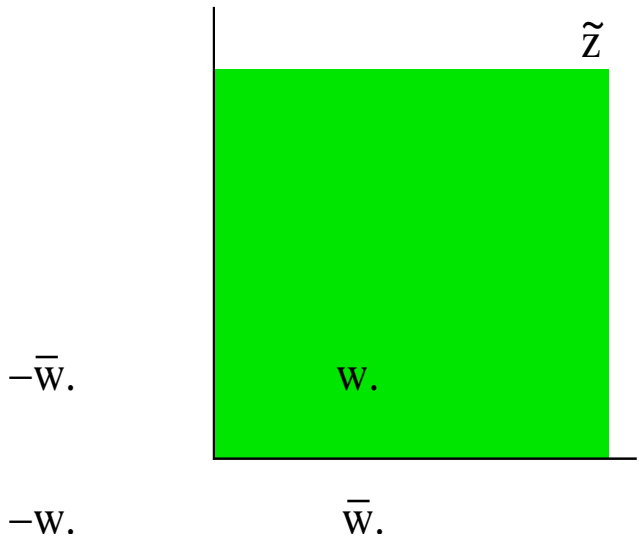}} \par}
\caption{intersections at right angles\label{90degrees}}
\end{figure}

Let us first consider string states at an intersection with branes at right
angles, in a pair of dimensions \( X_{1} \), \( X_{2} \). The set
up is as shown in figure~\ref{90degrees}.
We will refer to the complex world sheet coordinate as \( \tilde{z} \)
and map the space-time coordinates onto it directly, \( X_{1}+iX_{2}=\tilde{z} \).
Consider the Green function at a point in the plane for \( X_{1} \).
To calculate this we can use the method of images as shown in the
figure. The boundary condition is Neumann on the \( X_{1} \) axis
and Dirichlet on the \( X_{2} \) axis, so to take account of this
we need 3 images with the Green functions of those at negative \( X_{2} \)
being added negatively. The total Green function is given by 
\begin{eqnarray}
G(\tilde{z},\tilde{w}) & = & G_{0}(\tilde{z},\tilde{w})+G_{0}(\tilde{z},\overline{\tilde{w}})-G_{0}(\tilde{z},-\overline{\tilde{w}})-G_{0}(\tilde{z},-\tilde{w})\\
 & = & -\frac{\alpha '}{2}\ln \frac{|\tilde{z}-\tilde{w}|}{|\tilde{z}+\tilde{w}|}\frac{|\tilde{z}-\overline{\tilde{w}}|}{|\tilde{z}+\overline{\tilde{w}}|}
\end{eqnarray}
Now map the world-sheet to the usual upper half plane by making the
conformal transformation \( z^{2}=\tilde{z} \). In these coordinates
the Green function becomes 
\begin{equation}
G(z,w)=-\frac{\alpha '}{2}\ln
\frac{|\sqrt{z}-\sqrt{w}|}{|\sqrt{z}+\sqrt{w}|}-\frac{\alpha '}{2}\ln
\frac{|\sqrt{z}-\sqrt{\overline{w}}|}{|\sqrt{z}+\sqrt{\overline{w}}|},
\end{equation}
which can be recognized as the Green function of the original point
in the upper half plane, plus its image at \( \overline{w} \) in
the lower half plane, in the presence of a \( Z_{2} \) twist operator,
\( \sigma _{+} \) , placed at the origin~\cite{mansfield,fairlie,hamidi}. Consequently vertex operators
involving states with ND boundary conditions must include such operators.
Note that if the string has a Neumann condition at both ends (e.g.
if only one end was attached to the \( X_{1} \) brane with the second
end free) we would add all the images and find the usual untwisted
Green function 
\begin{equation}
G(z,w)=-\frac{\alpha '}{2}\ln |z-w|-\frac{\alpha '}{2}\ln |z-\overline{w}|.
\end{equation}

\begin{figure}
{\centering \resizebox*{0.6\textwidth}{0.4\textheight}{\includegraphics{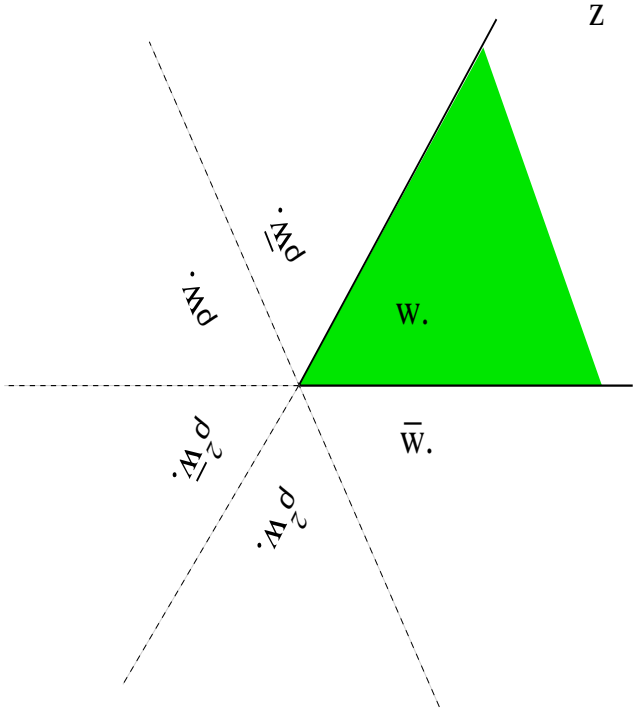}} \par}

\caption{The images for branes at \protect\( \pi /3\protect \), where \protect\( \rho =e^{2\pi i/3}\protect \).\label{60}}
\end{figure}

For more general angles one expects the appropriate twisted Greens
functions to appear. Indeed, writing the complex boson as \( X=X_{1}+iX_{2} \),
we can proceed in a similar manner, as shown in figure~\ref{60} for
\( \pi /3 \). 
The total contribution to the Green function is given by the sum of
the images;
\begin{equation}
G(\tilde{z},\tilde{w})=\sum _{l=0}^{m-1}\, \rho ^{l}G_{0}(\tilde{z},\rho
^{l}\tilde{w})+\rho ^{l}G_{0}(\tilde{z},\rho ^{l}\overline{\tilde{w}})
\end{equation}
where \( \rho =e^{\frac{2\pi il}{m}} \) and the intersection angle
is taken to be \( \pi /m \). Again transforming to the upper half
plane with \( z=\tilde{z}^{m} \) we find
\begin{equation}
G(z,w)=\sum _{l=0}^{m-1}\, \rho ^{l}G_{0}(z^{\frac{1}{m}},\rho
^{l}w^{\frac{1}{m}})+\rho ^{l}G_{0}(z^{\frac{1}{m}},\rho
^{l}\overline{w}^{\frac{1}{m}})
\end{equation}
(where the \( \frac{1}{m} \) roots are the trivial ones) which we
recognize as the Green function of the original point in the upper
half plane, plus its image at \( \overline{w} \) in the lower half
plane, in the presence of a \( Z_{m} \) twist operator, \( \sigma _{+} \)
, at the origin~\cite{hamidi}. 

\begin{figure}
\begin{center}
 \epsfig{file=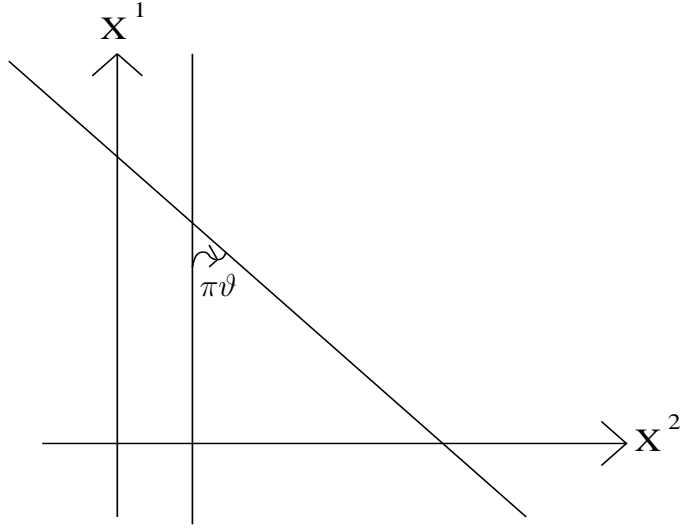,height=70mm,width=90mm}
\put(-183,87){\large{$\pi\vartheta$}}
\caption{A `twisted' open string state}
\label{stringpic}
\end{center}
\end{figure}

This is of course entirely consistent with the fractional mode expansion of
string states stretched between branes. In particular, consider a string
stretched between two D-branes intersecting at an angle $\pi \vartheta$, as depicted in figure~\ref{stringpic}. The boundary conditions are,
\begin{equation}
\label{bc}
\begin{array}{l}
\partial_{\tau}X^{2}(0)=\partial_{\sigma}X^{1}(0)=0, \\
\partial_{\tau}X^{1}(\pi)+\partial_{\tau}X^{2}(\pi)\cot(\pi\vartheta)=0, \\
\partial_{\sigma}X^{2}(\pi)-\partial_{\sigma}X^{1}(\pi)\cot(\pi\vartheta)=0,
\end{array}
\end{equation}
which determine the holomorphic solutions to the string equation of motion to be,
\begin{equation}
\label{modeexp}
\begin{array}{l}
\partial X(z)=\sum_{k}\alpha_{k-\vartheta}z^{-k+\vartheta-1},\\
\partial \bar{X}(z)=\sum_{k}\bar{\alpha}_{k+\vartheta}z^{-k-\vartheta-1}.
\end{array}
\end{equation}
We can define the  worldsheet coordinate, $z=-e^{\tau-i\sigma}$, which has domain the upper-half
complex plane. This can be extended to the entire complex plane using the
`doubling trick', i.e. we define,
\begin{equation}
\partial X(z)= \left\{ \begin{array}{ll}
                         \partial X(z) & \mbox{Im}(z) \geq 0 \\
                         \bar{\partial}\bar{X}(\bar{z}) & \mbox{Im}(z) <
                         0
                        \end{array} \right.,
\end{equation}
and similarly for $\partial \bar{X}(z)$.

The mode expansion in~(\ref{modeexp}) is identical to that of a
closed string state in the presence of a $\mathbb{Z}_{N}$ orbifold twist
field (with the replacement $\frac{1}{N}=\vartheta$). Thus
an open string stretched between intersecting D-branes is analogous
to a twisted closed string state on an orbifold and to take account of 
this we introduce a twist field $\sigma_\vartheta
(w,\bar{w})$ that changes the boundary conditions of
$X$ to be those of eq.(\ref{bc}), 
where the intersection point of the two D-branes
is at $X(w,\bar{w})$. Then, in an identical  manner to the closed string
case, we obtain the OPEs,
\begin{equation}
\label{opes}
\begin{array}{l}
\partial X(z) \sigma_{\vartheta}(w,\bar{w}) \sim
(z-w)^{-(1-\vartheta)}\tau_{\vartheta}(w,\bar{w}), \\
\partial \bar{X}(z) \sigma_{\vartheta}(w,\bar{w}) \sim
(z-w)^{-\vartheta}\tau'_{\vartheta}(w,\bar{w}),
\end{array}
\end{equation}
where $\tau'_{\vartheta}$ and $\tau_{\vartheta}$ are excited twists. Also,
the local monodromy conditions for transportion
around $\sigma_{\vartheta}(w,\bar{w})$ are,
\begin{equation}
\label{mono}
\begin{array}{l}
\partial X(e^{2 \pi i}(z-w))=e^{2\pi i\vartheta}\partial X(z-w), \\
\partial \bar{X}(e^{2 \pi i}(z-w))=e^{-2\pi i\vartheta }\partial \bar{X}(z-w).
\end{array}
\end{equation}
The mode expansion for $X$ is then,
\begin{equation}
X(z,\bar{z})=\sqrt{\frac{\alpha'}{2}}\sum_{k}\left( \frac{\alpha_{k-\vartheta}}{k-\vartheta}z^{-k+\vartheta}+\frac{\tilde{\alpha}_{k+\vartheta}}{k+\vartheta}\bar{z}^{-k-\vartheta}\right),
\end{equation}
with the right and left moving modes being mapped into upper and lower
half planes. A similar mode expansion is obtained for the fermions with
the obvious addition of \( \frac{1}{2} \) to the boundary conditions
for NS sectors.  

\section{Three point functions}
\begin{figure}
\begin{center}
 \epsfig{file=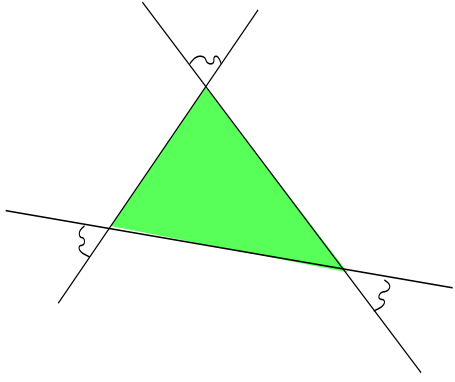,height=50mm,width=60mm}
\caption{three point interaction}
\end{center}
\label{IB}
\end{figure}

Now we proceed to calculate the classical part of the three point function, which in
particular includes the yukawa interactions. We will consider the case
of a compactified space which is factorizable into 2 tori, $T_{2} \times T_{2}
\times T_{2}$, and in which branes $A,B,C$ wrap one cycles 
$L_{A,B,C}$. As shown in figure~\ref{IB}, the string states are localised at
the vertices of a triangle whose boundary consists of the D-branes which wrap
the internal space. One would expect the amplitude to be dominated by an
instanton, and to be 
proportional to $e^{-\frac{1}{2\pi \alpha}A}$ where $A$ is the area of the
triangle worldsheet 
(also, due to the toroidal geometry we would expect a contribution
from wrapped triangles). This expectation is born out in the 
following calculation.

Denote the spacetime coordinates of our torus subfactor by $X=X^{1}+iX^{2}$
and $ \bar{X}=X^{1}-iX^{2}$. The bosonic
field $X$ can be split up into a classical piece, $X_{cl}$, and a quantum
fluctuation, $X_{qu}$. The amplitude then 
factorises into classical and quantum contributions,
\begin{equation}
Z=\sum_{\langle X_{cl} \rangle}e^{-S_{cl}}Z_{qu}.
\end{equation}
$X_{cl}$ must satisfy the string equation of motion and possess the correct
asymptotic behaviour near the triangle vertices. 

The three point function requires 3 twist vertices,
$\sigma_{\vartheta_{i}}(z_{i},\bar{z}_{i})$, corresponding to the three relevant
intersections of the D-branes. These vertices are attached to the boundary of
the tree-level disc diagram which can be conformally mapped to the upper-half
plane. Then our classical solution $\partial X_{cl}$ is determined, up to a
constant, by its holomorphicity and asymptotic 
behaviour at each D-brane intersection,
which is given by the OPEs in eq.(\ref{opes}). 
Hence, mapping the spacetime coordinates $X$ and $\bar{X}$
into the worldsheet coordinate z, we obtain,
\begin{equation}
\begin{array}{ll}
\partial X_{cl}(z)= a\omega(z), & \partial \bar{X}_{cl}(z)=\bar{a}
\omega'(z), \\
\bar{\partial} X_{cl}(\bar{z})= b\bar{\omega}'(\bar{z}), & \bar{\partial} \bar{X}_{cl}(\bar{z})=\bar{b}\bar{\omega}(\bar{z}),
\end{array}
\end{equation}
where,
\begin{equation}
\begin{array}{ll}
\omega(z)=\prod_{i=1}^{3} (z-z_{i})^{-(1-\vartheta_{i})} &
\omega'(z)=\prod_{i=1}^{3} (z-z_{i})^{-\vartheta_{i}},
\end{array}
\end{equation}
and $a,\bar{a},b,\bar{b}$ are normalisation constants. Here we have used the
doubling trick which also requires that $a^{*}=\bar{b}$ and $\bar{a}=b^{*}$ (upto a phase). We can therefore write,
\begin{equation}
\label{action}
S_{cl}=\frac{1}{4 \pi \alpha}\left( |a|^{2}\int d^{2}z |\omega(z)|^{2}
  +|\bar{a}|^{2}\int d^{2}z |\omega'(z)|^{2} \right),
\end{equation}
The contribution to $S_{cl}$ from $|\omega'(z)|$ diverges, hence we set $\bar{a}=0$.

\begin{figure}
\centering
\epsfig{file=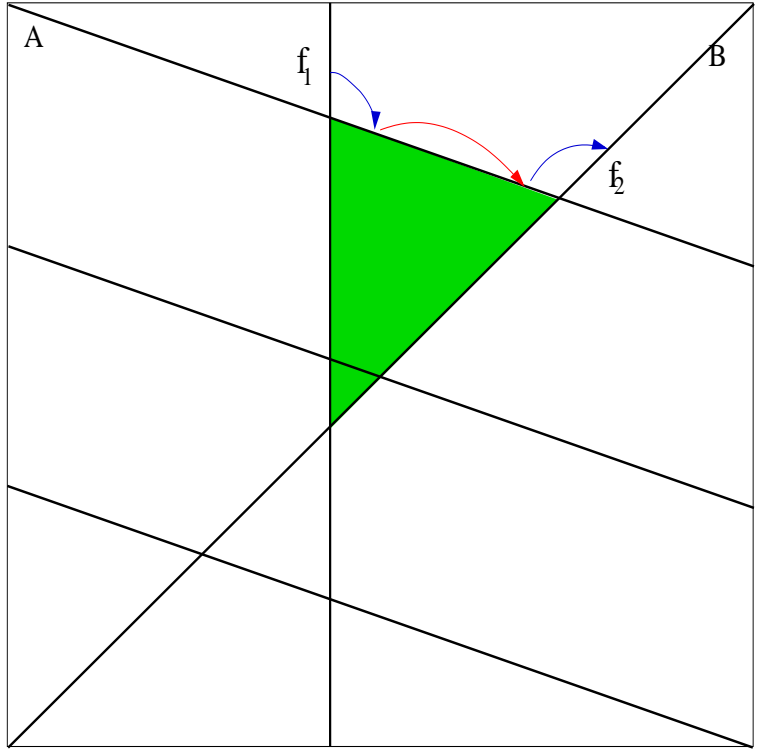, height=100mm,width=100mm}
\caption{Transporting $X(z,\bar{z})$ around two twist fields}
\label{target}
\end{figure}

The normalisation constants are determined from the global
monodromy conditions, i.e. the transformation of $X$ as it is transported
around more than one twist operator, such that the net twist is zero. We determine this from the action of a
single twist operator, $\sigma_{\vartheta}(w,\bar{w})$. In the 
closed string case this field acts to rotate and shift $X(z,\bar{z})$;
\begin{equation}
X(e^{2\pi i}z,e^{-2\pi i}\overline{z})=\theta X(z,\overline{z})+v
\end{equation}
where \( \theta  \) is a complex phase factor
with phase \( \pi \vartheta  \), and \( v \) is over a coset of
the  toroidal compactification lattice \( \Lambda  \) which depends on the
fixed point \( f \); 
\begin{equation}
v=(1-\theta )(f+\Lambda ).
\end{equation}
Ignoring the lattice for the moment, the equivalent statement for open strings at intersections is that 
\begin{equation}
\label{trans}
X(e^{2\pi i}z,e^{-2\pi i}\overline{z})= e^{2\pi i\vartheta}X +(1-e^{2\pi i\vartheta})f,
\end{equation}
where f is the intersection point of the two D-branes. This can be seen from
the local monodromy conditions~(\ref{mono}) and the fact that $f$ must be left
invariant. The global monodromy of $X$ is then simply a product of such actions.
On integrating around 2 twist fields, the strings are embedded in the target
space as shown in figure~\ref{target}. The portion of integration around each
vertex takes $X(z,\bar{z})$ from one brane to another, while integrating
between two vertices introduces a shift along that particular brane.

\begin{figure}
\centering
\epsfig{file=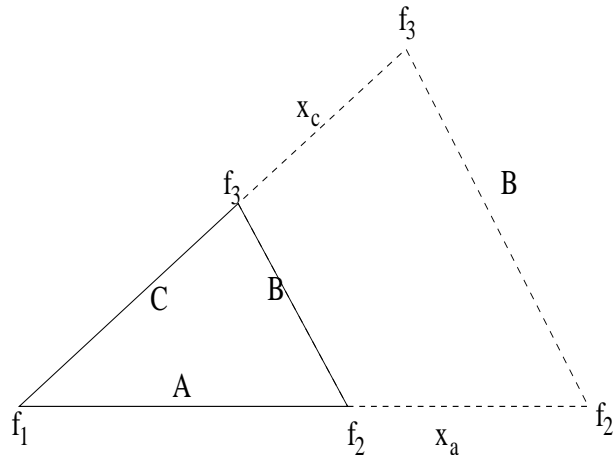, height=60mm,width=80mm}
\caption{Wrapping triangles}
\label{triangle}
\end{figure}

To determine the shifts, we must consider triangles that wrap the torus and
whose vertices are at the same intersection points. (Other shifts would contribute to other three point functions.)  
Keeping $f_{1}$ fixed and extending the
triangle as shown in figure~\ref{triangle}, we obtain a contribution to the
same three point function provided, 
\begin{equation}
\begin{array}{ll}
x_{A}=k_{A}L_{AB}|I_{AB}| & x_{C}=k_{A}L_{CB}|I_{CB}|,
\end{array}
\end{equation}
where $k_{A},k_{B} \in \mathbb{Z}$, $I_{ij}$ is the intersection number of
branes $i$ and $j$ and $L_{ij}$ is the displacement between
successive $i,j$ intersections along the $i^{th}$ brane. 
Using congruency of the triangle we obtain,
\begin{equation}
\begin{array}{ll}
k_{A}=l \frac{|I_{CB}|}{gcd(|I_{CB}|,|I_{AB}|)} & k_{C}=l
\frac{|I_{AB}|}{gcd(|I_{CB}|,|I_{AB}|)},
\end{array}
\end{equation}
where $l \in \mathbb{Z}$. On the other hand, if we instead keep $f_{2}$ fixed
we obtain,
\begin{equation}
\begin{array}{ll}
k_{A}=l' \frac{|I_{CB}|}{gcd(|I_{CB}|,|I_{AC}|)} & k_{B}=l'
\frac{|I_{AC}|}{gcd(|I_{CB}|,|I_{AC}|)},
\end{array}
\end{equation}
where $l'$ is a different integer. 
Hence our lattice shifts must be of the form,
\begin{equation}
x_{A}(l)=l\frac{|I_{CB}| L_{A}}{gcd(|I_{CB}|,|I_{AB}|,|I_{AB}|)},
\end{equation}
and similar for $x_{B,C}$ with the obvious replacment of indices, but with the 
same $l$. This is similar to the case discussed in ref.\cite{yukawa}.

\begin{figure}[h]
\centering
\epsfig{file=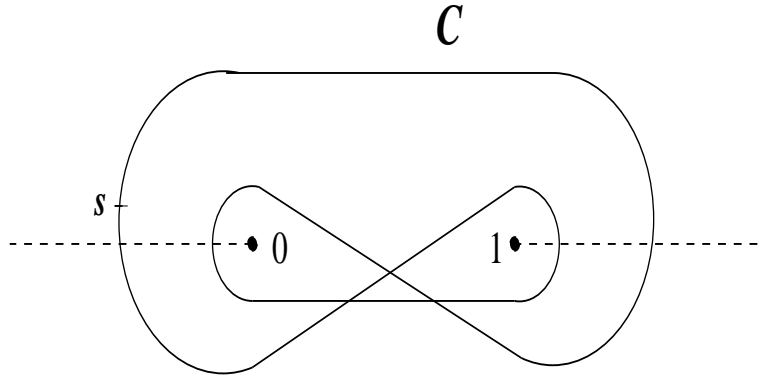, height=50mm, width=100mm}
\caption{Closed curve with net twist zero required for global monodromy condition.}
\label{curve}
\end{figure}

Considering all closed curves,
$C_{i}$, with net twist zero, we obtain the global monodromy conditions,
\begin{equation} 
\label{globmono}
\Delta _{C_{i}}X=\oint _{C_{i}}dz\partial X(z)+\oint
_{C_{i}}d\overline{z}\bar{\partial}X(\bar{z})=v_{i},
\end{equation}
In our case there is only one independent, net twist zero, closed curve. This
is shown in figure~\ref{curve}. We have set $z_{1}=0, z_{2}=1$ and
$z_{3} \rightarrow \infty$ using $SL(2,\mathbb{R})$ invariance and the dashed
lines denote branch cuts. Evaluating the contour integral we get,
\begin{equation}
\label{B}
\oint_{C} dz \omega(z)=-(-z_{\infty})^{-(1-\vartheta_{3})}4\sin
  \pi \vartheta_{2}\sin \pi\vartheta_{1} e^{\pi i \vartheta_{1}}
  \frac{\Gamma(\vartheta_{1})\Gamma(\vartheta_{2})}{\Gamma(1-\vartheta_{3})} ,
\end{equation} 
where phases from branch cuts are determined by starting at S, which we
assume is in the principal branch of each factor in the integrand. The
global monodromy condition~(\ref{globmono}) is independent of this convention provided we
order the transformations~(\ref{trans}) according to the path taken around
the curve C, again starting at S. This gives,
\begin{equation}
\label{A}
\Delta_{C}X_{cl}=4 \sin \pi \vartheta_{1} \sin \pi \vartheta_{2} e^{-i(\vartheta_{1}-\vartheta_{2})}(f_{1}-f_{2}+x_{A}(l)),
\end{equation}
substituting~(\ref{A}) and~(\ref{B}) into~(\ref{globmono}) we then
determine,
\begin{equation}
\label{a}
|a|^{2}=|-z_{\infty}|^{2(1-\vartheta_{3})}|f_{1}-f_{2}+x_{A}(l)|^{2}\frac{\Gamma(1-\vartheta_{3})}{\Gamma(\vartheta_{1})\Gamma(\vartheta_{2})}.
\end{equation}
We also require the integral,
\begin{equation}
\label{integral}
\int d^{2}z|\omega(z)|^{2}=|-z_{\infty}|^{-2(1-\vartheta_{3})}\sin \pi\vartheta_{2} (\Gamma(\vartheta_{2}))^{2}\frac{\Gamma(\vartheta_{3})\Gamma(\vartheta_{1})}{\Gamma(1-\vartheta_{3})\Gamma(1-\vartheta_{1})},
\end{equation}
 which can be performed using the method of~\cite{kawai} to relate open and
 closed string amplitudes. Finally, substituting~(\ref{integral})
 and~(\ref{a}) into~(\ref{action}) we obtain, 
\begin{equation}
S_{cl}=\sum_{l \in \mathbb{Z}}\frac{1}{2 \pi \alpha'}\left( \frac{\sin
      \pi \vartheta_{1} \sin \pi \vartheta_{2}}{2 \sin \pi \vartheta_{3}}|f_{1}-f_{2}+x_{A}(l)|^{2}\right).
\end{equation}
This is, as expected, the area of the triangle (and wrappings) defined by the intersecting D-branes that is swept out by the string worldsheet.

\section{Four point functions}
Having confirmed the efficacy of the conformal field theory techniques, 
let us now turn to the four point function on two sets of branes 
intersecting at angles. Processes involving contact interactions between
4 fermions have previously only been considered for orthogonal D branes
in ref.\cite{benakli}. For theories with branes at angles
these processes are particularly important because the sector of chiral
fermions is rather independent of the general set up, whereas the
scalars are more model dependent and may be tachyonic. As for the
orthogonal case, we shall see that the processes get contributions
from intermediate Kaluza-Klein excitations, winding modes and string
excitations. 

The massless fermions of interest appear in the Ramond
sector with charges, \( q_{i=0..3} \) for the 4 complex transverse
fermionic degrees of freedom given by one of the following; 
\begin{equation}
\begin{array}{lll}
q & = & (+\frac{1}{2},\, \, \vartheta _{1}-\frac{1}{2},\vartheta _{2}-\frac{1}{2},\vartheta _{3}-\frac{1}{2})\\
q & = & (\pm \frac{1}{2},\pm \frac{1}{2},\, \, \vartheta _{2}-\frac{1}{2},\vartheta _{3}-\frac{1}{2})\\
q & = & (\pm \frac{1}{2},\pm \frac{1}{2},\pm \frac{1}{2},\, \, \vartheta _{3}-\frac{1}{2})
\end{array}
\end{equation}
depending on the type of intersection. For example \( D_{6} \) branes
intersecting in \( (T_{2})^{3} \) are of the first kind. The GSO
projection leaves only one 4D spinor, and the theory is chiral. For
special values of angles (0 for example) supersymmetry may be restored,
but generally supersymmetry is completely broken, and the scalars
can be heavy or tachyonic. Fermions for \( D5 \) branes intersecting
in \( (T_{2})^{2}\times C/Z_{N} \) are of the second kind. Initially
the GSO projection will leave only half the space time spinor degrees
of freedom leaving a non-chiral theory. Hence a further orbifolding
on the 1st complex dimension is required to get a chiral theory. Finally
\( D_{4} \) branes intersecting in \( T_{2}\times C_{2}/Z_{N} \)
correspond to the last choice. Again the GSO projection leaves 4 states
which need to be further projected out by orbifolding in the \( C_{2} \)
dimensions. The particular orbifoldings do not effect the Ramond charges
above so the quantum part of the amplitude will be unaffected by it.
The classical part of the amplitude depends purely on the world sheet
areas. However as the orbifolding is orthogonal to the space in which
the branes are wrapping, it will not affect the classical part either.
The only affect of the orbifolding is therefore in projecting
out the chiralitites above. In addition we shall not consider orientifolding 
which does not effect the quantum part of the inetraction, except in so 
far as it determines the gauge groups. For the classical part, one would 
expect complication due to the presence of image branes, but these could be 
treated as essentially separate branes for the present discussion. Hence 
our results can be easily adapted to those cases as well. 

The four fermion scattering amplitude is given by a disc diagram with
4 fermionic vertex operators in the $-1/2$ picture, 
\( V^{(a)} \) on the boundary. The diagram is
then mapped to the upper half plane with vertices on the boundary.
The positions of the vertices ($x_1\ldots x_4$) will eventually be 
fixed by $SL(2,\mathbb{R})$ invariance
to \( 0,x,1,\infty  \) (where \( x \) is real), so that
the 4 point ordered amplitude can be written 
\begin{equation}
(2\pi )^{4}\delta ^{4}(\sum _{a}k_{a})\,
A(1,2,3,4)=\frac{-i}{g_{s}l_{s}^{4}}\int ^{1}_{0}dx\langle
V^{(1)}(0,k_{1})V^{(2)}(x,k_{2})V^{(3)}(1,k_{3})V^{(4)}
(\infty ,k_{4})\rangle .
\end{equation}
To get the total amplitude we sum over all possible orderings;
\begin{eqnarray}
A_{total}(1,2,3,4)&=&A(1,2,3,4)+A(1,3,2,4)+A(1,2,4,3) \nonumber \\
&+& A(4,3,2,1)+A(4,2,3,1)+A(4,3,1,2).
\end{eqnarray}
The vertex operators for the fermions are of the form 
\begin{equation}
V^{(a)}(x_{a},k_{a})=const\, \lambda ^{a}\, u_{\alpha }S^{\alpha }\prod
_{i}\sigma_{\vartheta_i}\, e^{-\phi /2}e^{ik_{a}.X}(x_{a}).
\end{equation}
Here \( u_{\alpha } \) is the space time spinor polarization, and
\( S^{\alpha } \) is the so called spin-twist operator of the form
\begin{equation}
S^{\alpha }=\prod _{i}:\exp (iq_{i}H_{i}):
\end{equation}
with conformal dimension 
\begin{equation}
h=\sum _{i}\frac{q_{i}^{2}}{2},
\end{equation}
and \( \sigma_{\vartheta_i} \) is the \( \vartheta  \) 
twist operator acting
on the \( i \)'th complex dimension, with conformal dimension
\begin{equation}
h_{i}=\frac{1}{2}\vartheta _{i}(1-\vartheta _{i}).
\end{equation}
The calculation of the 4-point function of the bosonic twist operators
can be done analogously to the closed string case~\cite{dixon}, 
with only minor modifications to take account of the image Green function and 
the fact that the vertices are on the real axis. 

For completeness we now outline the derivation. Consider the contribution
from a single complex dimension in which the branes intersect with angle $\vartheta \pi$. We begin with the asymptotic behaviour of the 
Green function in the vicinity of the twist operators. As we saw earlier, 
we take account of the world-sheet boundary by adding an image charge.
The Green function can then be written
\begin{equation}
G(z,w;z_i)=g(z,w;z_i)+g(z,\overline{w};z_i)
\end{equation}
where $g(z,w;z_i)$ is the usual Green function for the 
closed string. It has the following asymptotics 
\begin{eqnarray}
g(z,w;z_i)
&\sim & \frac{1}{(z-w)^2}+finite \hspace{0.7cm}  z\rightarrow w 
\nonumber \\ 
&\sim & \frac{1}{(z-x_{1,3})^{-\vartheta}} \hspace{1.6cm}  z\rightarrow x_{1,3} 
\nonumber \\ 
&\sim & \frac{1}{(z-x_{2,4})^{-(1-\vartheta )}} \hspace{1cm}  z\rightarrow x_{2,4} 
\nonumber \\ 
&\sim & \frac{1}{(w-x_{1,3})^{-(1-\vartheta)}} \hspace{1cm}  w\rightarrow x_{1,3} 
\nonumber \\ 
&\sim & \frac{1}{(w-x_{2,4})^{-\vartheta}} \hspace{1.6cm}  w\rightarrow x_{2,4} 
\nonumber \\ 
\end{eqnarray}
and as we have seen the holomorphic fields are 
\begin{eqnarray}
\partial X(z) & \equiv & \omega_\vartheta(z) =
[(z-x_1)(z-x_3)]^{-\vartheta}
[(z-x_2)(z-x_4)]^{-(1-\vartheta)}
\nonumber \\
\partial \overline{X}(z) & \equiv & \omega_{1-\vartheta}(z) =
[(z-x_1)(z-x_3)]^{-(1-\vartheta)}
[(z-x_2)(z-x_4)]^{-\vartheta}.
\end{eqnarray}
This half of the Green 
function may now be determined upto an additional term usually
denoted $A$ by inspection;
\begin{eqnarray}
g(z,w;z_i) &=& \omega_\vartheta(z)\omega_{1-\vartheta}(w)
\left\{ \vartheta 
\frac{(z-x_1)(z-x_3)
(w-x_2)(w-x_4) }{(z-w)^2} \right. \nonumber \\
&&
+ \left. (1-\vartheta)
\frac{(z-x_2)(z-x_4)
(w-x_1)(w-x_3) }{(z-w)^2}+A
\right\}.
\end{eqnarray}
Next one considers 
\begin{eqnarray}
\frac
{\langle T(z)\sigma_-\sigma_+\sigma_-\sigma_+ \rangle}
{\langle \sigma_-\sigma_+\sigma_-\sigma_+ \rangle}
&=&
\lim_{w\rightarrow z}[g_(z,w)-\frac{1}{(z-w)^2}] \nonumber \\
&=& \frac{1}{2}\vartheta (1-\vartheta)
\left( 
(z-x_1)^{-1}
+(z-x_3)^{-1}
-(z-x_2)^{-1}
-(z-x_4)^{-1}\right)^2\nonumber \\
&&
+\frac{A}{(z-x_1)
(z-x_2)
(z-x_3)
(z-x_4)}
\end{eqnarray}
where for shorthand we denote $\sigma_\pm \equiv \sigma_{\vartheta^{\pm 1}}$,
and compares it to the OPE of $T(z)$ with the twist operator
\begin{equation}
T(z)\sigma_+(x_2) \sim \frac{h_i}{(z-x_2)^2}+\frac{\partial_{x_2}\sigma_+(x_2)}{(z-x_2)}+\ldots
\end{equation}
Equating coefficients of $(z-x_2)^{-1}$, and then using $SL(2,R)$ invariance 
to fix $(x_1,x_2,x_3,x_4)= (0,x,1,x_\infty)$ yields a differential 
equation for ${\langle \sigma_-\sigma_+\sigma_-\sigma_+ \rangle}$ of the form 
\begin{equation}
\label{differ}
\partial_x \ln \left[ 
x^{\vartheta(1-\vartheta)}_\infty
{\langle \sigma_-\sigma_+\sigma_-\sigma_+ \rangle}
      \right]
=
\partial_x \ln \left[ (x(1-x))^{-\vartheta(1-\vartheta)} \right] 
-\frac{A(x)}{x(1-x)}
\end{equation}
where 
\begin{equation}
A(x)=-x_\infty^{-1} A(0,x,1,x_\infty).
\end{equation}
All that remains is to determine $A(x)$ which can be done using
monodromy conditions for $\partial_z X \partial_w \overline{X}$.
We proceed along the same lines as in the three point calculation and consider
the two independent loops \( C_{i} \) around combinations of twists that add up
to zero (again using the doubling trick), 
applying the global monodromy conditions~(\ref{globmono}).
On encircling a twist \( \sigma ^{+}_{f_{1}} \) associated with
states at intersection \( f_{1} \), plus antitwist, \( \sigma ^{-}_{f_{2}} \),
at intersection \( f_{2} \) the quantum part of $\partial X$ should be 
invariant. There are two independent pairs of twists and integrating 
around the the corresponding closed loops yields
\begin{equation}
A(x)=\frac{x(1-x)}{2}\partial_x \ln \left[ F(x)F(1-x) \right]
\end{equation}
where \( F(x) \) is the hypergeometric function
\begin{equation}
F(x)=F(\vartheta ,1-\vartheta ;1;x)=\frac{1}{\pi }\sin (\vartheta \pi )\int
^{1}_{0}dy\, y^{-\vartheta }(1-y)^{-(1-\vartheta )}(1-xy)^{-\vartheta }
\end{equation}
Inserting this into eq.(\ref{differ}), 
we find a contribution from the product of \( \vartheta_i  \)
twisted bosons of 
\begin{equation}
\prod_i \langle \sigma _{+}(x_{\infty })\sigma _{-}(1)\sigma _{+}(x)\sigma
_{-}(0)\rangle =const\, \frac{(x_{\infty }x(1-x))^{-\vartheta .(1-\vartheta
    )}}{[F(1-x).F(x)]^{1/2}}.
\end{equation}
In this expression we use a dot to indicate the product of the contributions from each complex dimension. 
When we collect all the contribution together the dependence on \( \vartheta _{i} \)
cancels between the bosonic twist fields and the spin-twist fields giving
\begin{equation}
\begin{array}{lll}
A(1,2,3,4) & = & -g_{s}\alpha '(\lambda ^{1}\lambda ^{2}\lambda ^{3}\lambda ^{4}+\lambda ^{4}\lambda ^{3}\lambda ^{2}\lambda ^{1})\, \int ^{1}_{0}dx\, x^{-1-\alpha 's}(1-x)^{-1-\alpha 't}\frac{1}{[F(1-x).F(x)]^{1/2}}\\
 &  & \times \left[ \overline{u}^{(1)}\gamma _{\mu }u^{(2)}\overline{u}^{(4)}\gamma ^{\mu }u^{(3)}\right]\sum e^{-S_{cl}(x)}
\end{array}
\end{equation}
where \( s=-(k_{1}+k_{2})^{2} \), \( t=-(k_{2}+k_{3})^{2} \), \( u=-(k_{1}+k_{3})^{2} \)
are the usual Mandlestam variables. We can check this expression against that 
in ref.\cite{benakli} where the branes considered 
were orthogonal D7 branes and D3 branes. 
The quantum part of the above expression with
intersections in two complex subplanes should agree with that from 
four ND fermions attached to two sets of coincident D3 
branes and a stack of D7 branes.
We find agreement upto a factor of $F(1-x)/F(x) $ in the integrand.

\begin{figure}
{\centering \resizebox*{0.8\textwidth}{0.4\textheight}{\includegraphics{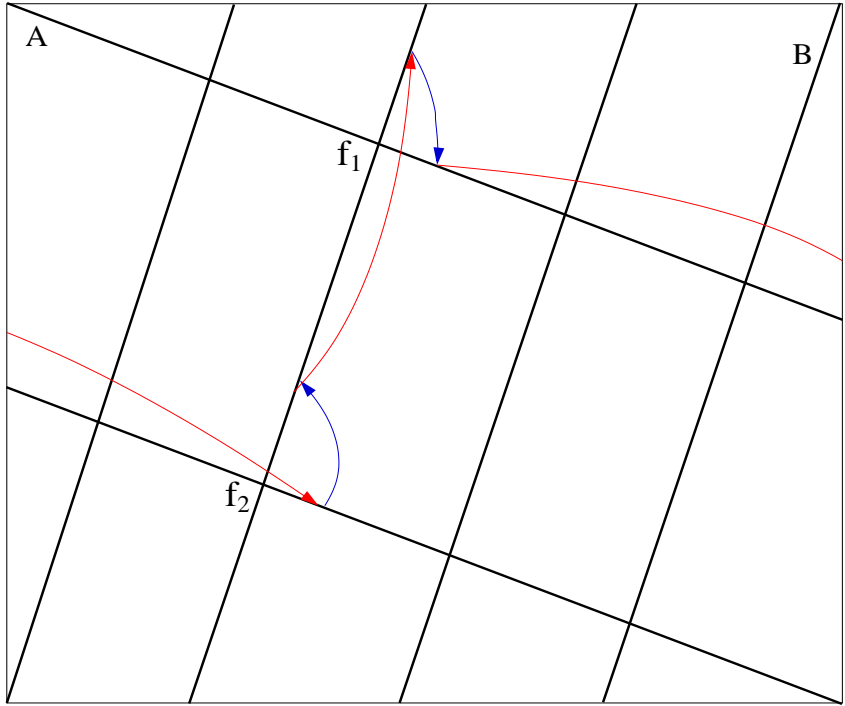}} \par}

\caption{Joined pairs of twisted/antitwisted states. \label{loops} }
\end{figure}

We now turn to the classical contribution to the four point function which 
can be calculated as for the three point function. On integrating around the 
closed loops we again can have arbitrary shifts $v_i$ along the D-branes.
Consider encircling a twist \( \sigma ^{+}_{f_{1}} \) associated with
states at intersection \( f_{1} \), plus antitwist, \( \sigma ^{-}_{f_{2}} \),
at intersection \( f_{2} \). The corresponding open strings can encircle
one of the branes appearing at the intersection 
determined by the relative positions of \( \sigma _{f_{1}}^{+} \)
and \( \sigma _{f_{2}}^{-} \) with respect to the branch cuts in
the field \( X \). The two possibilities are shown in figure~\ref{loops}.
A twisted state at \( f_{2} \) followed by an anti-twisted state
at \( f_{1} \) corresponds to a shift \( (1-\theta )(f_{1}-f_{2}+\Lambda _{A}) \)
where \( \Lambda _{A} \) is an integer number of shifts along the
\( A \)-brane. Note that $f_{1}-f_{2}$ is in this case a vector along the 
$A$ brane as well. An anti-twisted state at \( f_{1} \) followed by
twisted state at \( f_{2} \) causes a shift \( (1-\theta )(f_{1}-f_{2}+\Lambda _{B}) \) where now \( \Lambda _{B} \) is an integer shift along the \( B \)
brane and $f_{1}-f_{2}$ is now the displacement between the points along the 
$B$-brane.
The vertex ordering in the disc diagram means that in the amplitude
the factor \( \sum e^{-S_{cl}} \) now splits into separate sums over
\( \Lambda _{A} \) or \( \Lambda _{B} \),\( \sum \equiv \sum _{\Lambda _{A},\Lambda _{B}} \). The classical action is then found to be 
\begin{equation}
S_{cl}=\frac{\sin \vartheta \pi }{4\pi \alpha '\tau }\left( |v'_{A}|^{2}+\tau
  ^{2}|v'_{B}|^{2}\right), 
\end{equation}
where in the sums over \( \Lambda _{A} \) we have 
\begin{equation}
v'_{A,B}=\Delta_{A,B}^i f+nL_{A,B},
\end{equation}
Here \( n\in Z \), and \( L_{A,B} \) are vectors in the two torus describing
the wrapped D-branes. We have defined $\Delta_{A,B}^{i} f$ to be the displacement 
between the intersections involved in the $C_i$ loop taken along the $A,B$ 
brane, and we include both combinations (i.e. $i=1,2$ going respectively 
with $A,B$ or $B,A$). In addition we have defined 
\( \tau (x)=\frac{F(1-x)}{F(x)} \)
for convenience. (For \( Z_{2} \) twists, i.e. intersections at right-angles,
this would be the modular parameter of a {}``fake'' annulus but
it has no such interpretation for more general intersections.) 

\begin{figure}
{\centering \resizebox*{0.8\textwidth}{0.4\textheight}{\includegraphics{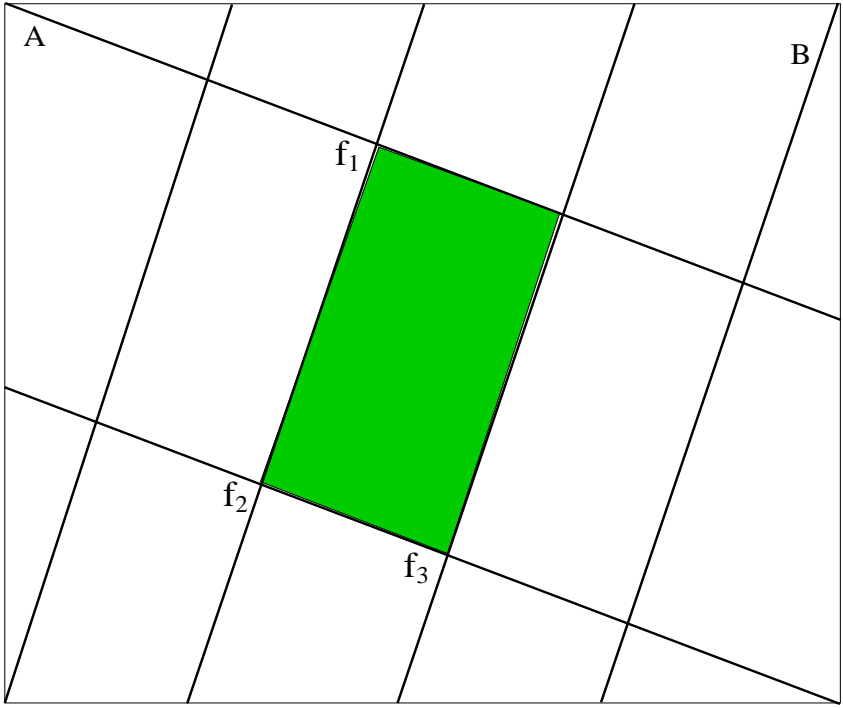}} \par}
\caption{The leading contribution to a generic four point diagram.\label{generic}}
\end{figure}

\begin{figure}
{\centering 
\resizebox*{0.8\textwidth}{0.4\textheight}
{\includegraphics{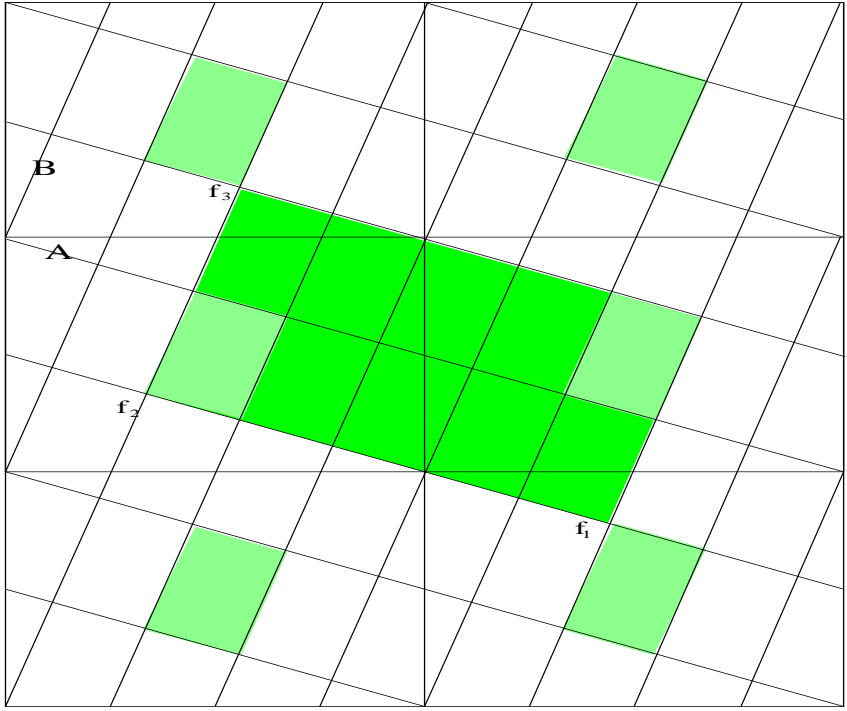}} \par}
\caption{The first 
subleading contribution to the same amplitude. \label{generic2} }
\end{figure}

To illustrate, consider the generic open string four point diagram, 
shown in fig~\ref{generic}.
In this case one expects the action to go as the area. To show this
we can use a saddle point approximation for the \( x \) integral
in \( A(1,2,3,4). \) This gives \( \tau (x_{s})=|v'_{A}|/|v'_{B}| \)
where, for strings stretched between \( f_{1} \) and \( f_{2} \)
propagating along the \( A \)-brane direction
we choose \( \Delta_A f = f_{3}-f_{2} \) and \( \Delta_B f=f_{1}-f_{2} \).
This is the leading contribution, but there 
is an additional contribution to the amplitude where the displacement is
\( \Delta_A f = f_{1}-f_{2} \) measured along the $A$-brane
and \( \Delta_B f =f_{2}-f_{3} \) measured along the $B$-brane. This 
corresponds to diagrams where
a string stretched between \( f_{2} \) and \( f_{3} \) propagates
in the \( f_{1}-f_{2} \) direction, as shown in fig.\ref{generic2}.

The accuracy of the saddle approximation
is a function of the width of the saddle, 
given by \( \sqrt{4\pi \alpha '/R_{c}}\sim \frac{l_{s}}{R_{c}} \),
where \( R_{c} \) is the compactification scale (\( R_{c}\sim L_{A},L_{B} \)).
Thus, as expected the approximation of world-sheet instanton suppression
breaks down when the size of the world sheet is comparable to the
D-brane thickness. 
Substituting back into the action we find 
\begin{equation}
A(1,2,3,4)\sim \sum _{\Lambda _{A},\Lambda _{B}}\, fermion-factors\times
\left( \frac{4\pi \alpha '}{R_{c}}\right) ^{2}\exp \left( -\frac{1}{2\pi
    \alpha '}\sin \vartheta \pi \, |v'_{A}||v'_{B}|\right) .
\end{equation}
 We find the expected suppression from the mass of the intermediate
stretched string state, times by an instanton suppression factor again 
given by the area of the world sheet. As pointed out in ref.\cite{ams}
diagrams such as this can lead to large flavour changing 
processes such as \( \tau \mu \rightarrow \mu e \) if the compactification 
scale is too small. 

Diagrams which do not
explicitly violate flavour, 
such as \( \mu ^{+}\mu ^{-}\rightarrow e^{+}e^{-} \) cannot be completely 
approximated this way.
For such processes the intersection separation in the leading term
\( v'_{B}=\Delta_B f_{ee} \) for one pair of twist operators is zero,
and so this term cannot be treated as above. Consider the summation
over \( \Lambda _{A} \) in \( v'_{A} \) for the other pair, whose
separation is \( v'_{A}=\Delta_A f_{e\mu }+nL_{A} \). Poisson resumming
we find 
\begin{equation}
\sum e^{-S_{cl}}=\sum _{p_{A}\in \Lambda _{A}^{*}}\sqrt{\frac{4\pi ^{2}\alpha
    '\tau }{L^{2}_{A}\sin \vartheta \pi }}\, \exp \left[ -\frac{4\pi
    ^{3}\alpha '\tau }{\sin \vartheta \pi }p_{A}^{2}\right] \, \exp \left[
  2\pi i\Delta f_{e\mu }.p_{A}\right] +subleading
\end{equation}
where \( p_{A}\in \Lambda ^{*}_{A} \)is summed over the dual lattice;
\begin{equation}
p_{A}=\frac{n_{A}}{|L_{A}|^{2}}L_{A}.
\end{equation}
This expression describes the leading exchange of gauge bosons plus
their KK modes along the \( A \) brane. (The subleading terms (those
with \( v'_{B}=n_{B}L_{B} \) with integer \( n_{B}\neq 0 \)) can
still be treated using the saddle point approximation above.)  
To obtain the field
theory result, we take the limit of coincident vertices; \( x\rightarrow 0 \)
or \( x\rightarrow 1 \). For example the former contribution can
be evaluated using the asymptotics 
\begin{equation}
F(x)\sim 1\, ,\, \, \, \, \, \tau \sim F(1-x)\sim \frac{1}{\pi }\sin
\vartheta \pi \, \ln \frac{\delta }{x}
\end{equation}
where \( \delta  \) is given by the digamma function \( \psi (z)=\Gamma '(z)/\Gamma (z); \)
\begin{equation}
\delta =\exp (2\psi (1)-\psi (\vartheta )-\psi (1-\vartheta )).
\end{equation}
We find \begin{eqnarray}
A(1,2,3,4) & = & g_{s}\left[ \overline{u}^{(1)}\gamma _{\mu }u^{(2)}\overline{u}^{(4)}\gamma ^{\mu }u^{(3)}\right] \frac{2\pi \sqrt{\alpha ' }}{L_{A}\sqrt{\sin \vartheta \pi} }\\
 &  & \times \, \left( \frac{1}{s}+2\sum ^{\infty }_{n=1}\frac{\cos \left( 2\pi \Delta f_{e\mu }.p_{n}\right) \, \delta ^{-\alpha 'M_{n}^{2}}}{s-M_{n}^{2}}\right) ,
\end{eqnarray}
where \( M_{n}=n/2\pi L_{A} \), and \( p_{n}=nL_{A}/|L_{A}|^{2} \).
This result agrees with a naive field theory calculation as shown in 
ref.\cite{ams}, provided that \( \alpha 'M_{1}^{2}\ll 1 \).
(That is, the brane separation should again be larger than the brane
thickness.) This can lead to severe flavour changing effects due
to the KK mode contribution which is flavour non-universal. 
We should note here that this is not in contradiction with the 
fact that there is an overall translational $U(1)$ 
symmetry on the torus, since the result is a function of the {\em relative} 
displacement. The leading term corresponds to massless gauge boson exchange so 
that we can normalize the coupling to the measured Yang-Mills couplings in 
the obvious way.

The extension to higher dimensional intersecting branes
follows straightforwardly, and we now find 
that the form factor \( \, \delta ^{-\alpha 'M_{n}^{2}} \)
naturally provides the UV cut-off which in the field theory has to
be added by hand. Physically the cut-off arises because the intersection
itself has thickness \( \sim \sqrt{\alpha '} \), and thus cannot
emit modes with a shorter wavelength. 

\section{Conclusion}

In summary, we have discussed the calculation of 
three point amplitudes for open strings
localised at D-brane intersections. We 
computed the four point ``contact interaction'' 
at tree level for diagrams involving fermions living at
two sets of intersecting D-branes. For the four point functions we 
were able to adapt the techniques of ref.\cite{dixon}
to the open string case, so that by considering the 
conformal field theory we obtained the required information to
compute the entire contact interaction. It receives contributions from both
KK modes and heavy string modes.
The obvious extension of this work 
would be to calculate the four 
point functions on three sets of intersecting branes. 
This would allow us to factorize the four point functions on the three 
point Yukawa couplings yielding additional information about the latter. 
In addition we considered only the simplest kind of toroidal 
compactification, and it would be interesting to contemplate the same 
process in more complicated set-ups, and also to apply it to a 
``realistic'' SM-like model. 

More generally these 
calculations may be used to discuss or support more generic field theoretic 
ideas in set-ups with fermions localized in extra dimensions. 
For example, in a parallel work \cite{ams}, it has been demonstrated 
in that very low (TeV) 
string scales are incompatible with the experimental absence of FCNCs. 
When the compactification scale is small we encounter subleading 
flavour changing diagrams involving stretched strings. When it is 
large it is instead the KK modes which contribute to FCNCs. 
Taken together these two 
contributions constrain the string scale. There
are many conceptual problems when discussing that type of effect 
that cannot be addressed in field
theory. For example, sums over Kaluza-Klein modes can diverge, so that
it is generally necessary to provide a UV cut-off. As is well known,
in the string calculation such a divergence should always be naturally 
regulated at the string scale, and this indeed happens automatically 
for the interactions discussed here. This has already been seen in the 
literature in for example \cite{ghilencea}. 

\bigskip 

\noindent {\bf Note Added} 

\noindent Whilst this paper was in the final stages of preparation we received 
ref.\cite{mirjam} which also discusses the calculation of 
both 3 and 4 point functions.

\section*{Acknowledgements}
It is a pleasure to thank David Fairlie, Paul Mansfield, Manel Masip and Jose
Santiago for discussions. We also thank Igor Klebanov for correcting an error
in an earlier version of this paper.
This work was funded by a PPARC studentship, and by Opportunity 
Grant PPA/T/S/1998/00833. 

\bibliography{CPbib}
\bibliographystyle{unsrt}
\end{document}